\documentclass[pra,twocolumn,showpacs]{revtex4}
\usepackage{graphicx,amsmath}

\begin{document}
\date{\today}

\title{Dynamics of double-well Bose-Einstein Condensates subject to external Gaussian white noise}

\author{Hanlei Zheng, Yajiang Hao and Qiang Gu\footnote[1]{Email: qgu@ustb.edu.cn}}

\address{Department of Physics, University of Science and Technology
  Beijing, Beijing 100083, China}

\date{\today}

\begin{abstract}
Dynamical properties of the Bose-Einstein condensate in double-well
potential subject to Gaussian white noise are investigated by
numerically solving the time-dependent Gross-Pitaevskii equation.
The Gaussian white noise is used to describe influence of the random
environmental disturbance on the double-well condensate. Dynamical
evolutions from three different initial states, the Josephson
oscillation state, the running phase and $\pi$-mode macroscopic
quantum self-trapping states are considered. It is shown that the
system is rather robust with respect to the weak noise whose
strength is small and change rate is high. If the evolution time is
sufficiently long, the weak noise will finally drive the system to
evolve from high energy states to low energy states, but in a manner
rather different from the energy-dissipation effect. In presence of
strong noise with either large strength or slow change rate, the
double-well condensate may exhibit very irregular dynamical
behaviors.

\end{abstract}


\maketitle

%%%%%%%%%%%%%%%%%%%%%%%%%%%%%%%%%%%%%%%%%%%%%%%%%%%%%%%%%%%%%%%%%%
\section{Introduction}

The atomic Bose-Einstein condensate (BEC) trapped in double-well
potentials builds up bosonic Josephson junction (BJJ)
\cite{Javanainen1986,Walls1996,Smerzi1997,Leggett1998,Collett1998}.
Since it exhibits abundant quantum properties in comparison to
condensate in a single trap, the double-well condensate has already been intensively investigated theoretically in the last few years. As a
BJJ, the double-well condensate can not only display dc, ac
Josephson effects and the Shapiro effect. It also exhibits an quantum
nonlinear effect, named the macroscopic quantum self-trapping (MQST)
\cite{Smerzi1997}. On the other hand, quantum fluctuation is
believed to give rise to fascinating influence on above dynamical
behaviors
\cite{Walls1997,Smerzi2000,Vardi2001,Cederbaum2009,Kroha2009,Vardi2009,Polls2010},
such as collapse and revival of quantum oscillations
\cite{Walls1997,Smerzi2000}, disappearance of coherence
\cite{Vardi2001,Cederbaum2009}, and destruction of the self-trapped
state \cite{Kroha2009}.

Experimentally the Josephson tunneling and MQST in a single BJJ have
been observed in 2005 \cite{Oberthaler2005}. Since then, more
progress has been made in studying static, thermal and dynamical
properties of the double-well condensate. Experimental investigation
of thermal induced phase fluctuations has been reported
\cite{Gati2006}. Measurements of the ac and dc Josephson effects in
BJJ have already been realized \cite{Levy2007}. The BJJ system has
also been used to perform interference-fringe experiments
\cite{Schmiedmayer2008} and to investigate the crossover from
Josephson dynamics to hydrodynamics \cite{Smerzi2011}. In above
theoretical studies, the double-well condensate is mainly treated as
an isolated system, but actually it is coupled to certain thermal
cloud and subject to environmental distortions in experiments.
Dissipation and noise effects play important roles in understanding
properties of BJJs.

The dissipative effect has been studied by several groups
\cite{Walls1998,Marino,gu12,Pitaev2001,Sols2003,Wimberger2008}. It
is suggested that the MQST state can be destroyed by energy
dissipation \cite{Walls1998,Marino,gu12}. Possible decoherence
caused by dissipation is also discussed \cite{Pitaev2001,Sols2003}.
It is also shown that dissipation could lead to enhancement of
coherence under specific conditions \cite{Wimberger2008}.

Here we consider a kind of noise effect on the double-well
condensate. Noise can be classified as ``internal noise'' or
``external noise'' with respect to its origins \cite{Sancho1982}.
Internal noise comes from the inside fluctuations of the system,
including quantum and thermal fluctuations, or from the exchange
symmetry of identical particles \cite{Gritsev2006,Anatoli2007}. On
the other hand, ``external noise'' is brought about by fluctuations
which are not ``self-originating''. It is induced by the
coupling between the system and its environment. For cold atomic
systems, external noise may originate from the magnetic field, laser
beams, or other externally applied random driving field. The present
work will focus on this kind of external noise.

%It is deserved to emphasize that the double-well BECs are not actually isolated, it is always coupled to the environment and the confinement are not homogeneous absolutely. And the condensate atoms inevitably interact with the surrounding thermal atoms.
%The noise effects will result in phase fluctuations, decoherence, dissipation and the shortening of lifetime are generally ignored for the weak fluctuation of condensates.

A number of theoretical works have been devoted to the influence of
external noise on the double-well condensate
\cite{Wimberger2008,Vardi2008,Ferrini2010}. The noise-induced
dephasing \cite{Vardi2008} and phase decoherence \cite{Ferrini2010}
have been predicted. These works based on the two-mode approximation mainly discussed the phase noise. The phase noise is introduced
by coupling a stochastic fluctuations either to the tunnel amplitude
or to the number-imbalance operator. In their treatment, spatial
information of the external noise is ignored.

The present paper deals with noise due to fluctuations of magnetic
field or the optical potential. We shall consider the time- and
space-dependent characteristics of the external noise by solving the
time-dependent Gross-Pitaevskii (GP) equation numerically. In
Sect.~\ref{sect:model}, we describe a noise model in which the
noise is simulated by spatially distributed stochastic potentials.
The concept of Gaussian white noise and the algorithm used to solve
the time-dependent GP equation for the double-well condensate are
briefly described. We present in Sect.~\ref{sect:diagram} the
obtained results and discuss the influence of noise on the dynamical
behaviors of system. A brief summary is given in the last section.

%%%%%%%%%%%%%%%%%%%%%%%%%%%%%%%%%%%%%%%%%%%%%%%%%%%%%%%%%%%%%%%%
\section{The model and method}
\label{sect:model}

We consider the BEC confined in a double-well potential is composed of atoms of mass $m$ with weakly
repulsive interaction. Dynamics of the double-well condensate is obtained by solving the
time-dependent GP equation numerically\cite{Bergeman2006, 2M}. We formulate the external
noise by an additional potential of stochastic strength fluctuation
in the space.

%\subsection{Gaussian white noises}

The external noise is modeled as the Gaussian white noise potential
$V_n(z,t)$, which is space- and time-dependent stochastically. So
$V_n(z,t)$ satisfies the following equations,
\begin{eqnarray}\label{NOISE1}
\langle V_{n}(z_i,t)\rangle=0,
\end{eqnarray}
and
\begin{eqnarray}\label{NOISE2}
\langle V_{n}(z_i,t)V_{n}(z_j,t^{\prime}) \rangle=2D_0\delta(t-t^{\prime})\delta(z_i-z_j),
\end{eqnarray}
where $\langle \rangle$ denotes averaging over both the space and
the time. The Dirac delta function in the correlation formula makes
sure $V_{n}(z,t)$ is ``white'' noise, $D_0$ is the fluctuation
amplitude of noise potential. Apparently, $D_0$
characterizes the strength or intensity of noise and it can be
controlled experimentally.

The noise correlation function in time at a given position is
defined as $g(t-t^{\prime})=\langle
V_{n}(z_i,t)V_{n}(z_i,t^{\prime})\rangle_T-\langle
V_{n}(z_i,t)\rangle_T^2 \sim \delta(t-t^{\prime})$ where $\langle
\rangle_T$ means that the average is just taken over time. Under
this definition, $\langle V_n(z,t)\rangle_T$ must be zero at all
given positions. In numerical calculations, $V_n(z,t)$ is produced
as a time-ordered series, $V_n(z,t_i)$, so the time average is
performed as $\langle V_{n}(z,t)\rangle_T= {\sum_{i=1}^M
V_{n}(z_i,t_i)\over M} $. $M$ is actually a finite number,
and $M\Delta t$ is the total time interval with $\Delta
t=t_{i+1}-t_i$ being the step interval of the noise potential series.
In our calculations, $M=100$ is already enough to
makes sure $\langle V_n(z,t)\rangle_T=0$. Therefore $\Delta t$ can
be used to evaluate the velocity of noise. Faster noise corresponds
to smaller $\Delta t$.
%\subsection{Time-dependent Gross-Pitaevskii  equation}

The general three dimensional time-dependent GP equation provides an
exact and fundamental description for our research, which can be
formulated as
\begin{equation}\label{GP3D}
i\hbar\frac{\partial\psi({\bf r};t)}{\partial t}=\left[-\frac{\hbar^2}{2m}\nabla^2+V({\bf r})+gN|\psi({\bf r};t)|^2\right]\psi({\bf r};t),
\end{equation}
where $\psi({\bf r};t)$ is the macroscopic wave function at position
${\bf r}$ and time $t$, $g=4\pi\hbar^2 a/m$ is the nonlinear
interaction with $a$ being the $s$-wave scattering length. The
external potential $V({\bf r};t)$ consists of the double well
$V_{dw}({\bf r})$ and the noise potential $V_n({\bf r};t)$.
$V_{dw}({\bf r})=\frac{m}{2}( \omega^{2}_x r_x^2 +
\omega^{2}_yr_y^2+\omega^{2}_z r_z^2) + V_{\rm
b}\mathrm{exp}(-r_z^2/q^2_0)$ fixes the double-well configuration,
where $\omega_{i} (i=x,y,z)$ is the trap frequencies in the
direction of $i$ \cite{Polls2010,Oberthaler2005} and $V_{\rm b}$ is
the barrier height. Experimentally, the Bose-Einstein condensates
with repulsive interaction in a Quasi-1D symmetric double-well
potential can be achieved by splitting one cigar-shaped atomic cloud
into two separated aligned cigars by a laser beam.

Set the double wells being along the $z$ direction and in $x$ and
$y$ directions the strong confinement is exerted
($\omega_x=\omega_y\gg \omega_z$). Thus the original three
dimensional condensates shall be reduced into a quasi one
dimensional cigar BECs and Eq. (\ref{GP3D}) is reduced into a
one-dimensional one. By scaling the length and energy as
$l_z=\sqrt{\hbar/(m\omega_z)}$ and $\hbar\omega_z$ respectively, the
dimensionless parameters are simplified as $z=r_z/l_z$,
$\tau=t\omega_z/2$ and
$\beta=\sqrt{\omega^{2}_x+\omega^{2}_y}/\omega_z$. Therefore, the
reduced one dimensional GP equation can be formulated as
\begin{eqnarray}\label{GP1D}
i\frac{\partial\bar\psi(z;\tau)}{\partial\tau}=\left[-\frac{\partial^2}{\partial z^2}+v_{dw}(z)+2v_{n}(z;t)+g_{1D}|\bar\psi(z;\tau)|^2\right]\bar\psi(z;\tau),
\end{eqnarray}
where $v_{dw}= z^2 + 2v_{\rm b}\mathrm{exp}(-z^2/q^{\prime 2}_0)$,
with $v_{dw}={V_{dw}}/({\hbar\omega_z})$,
$v_{n}={V_{n}}/({\hbar\omega_z})$ and $q^{\prime}_0=q_0/l_z$. We
obtain the dimensionless interaction parameter $g_{1D}=gNm\beta/(\pi
l_z \hbar^2)$ with $N$ the total atom number.

The wave function $ \bar\psi(z;\tau)$ satisfies the normalization
condition $\int^{\infty}_{-\infty}\mathrm{d}z|\bar\psi(z;\tau)|^2=1$. The fraction of atom number in the left
and right well are
$n_L(\tau)=\int^0_{-\infty}\mathrm{d}z|\bar\psi(z;\tau)|^2$
and
$n_R(\tau)=\int_0^{\infty}\mathrm{d}z|\bar\psi(z;\tau)|^2$,
respectively \cite{2M}. $\theta_L(\tau) = \mathrm{arctan}
\frac{\int^0_{-\infty}\mathrm{d}z \mathrm{Im}
[\bar\psi(z;\tau)]\rho(z;\tau)} {\int^0_{-\infty}\mathrm{d}z
\mathrm{Re}[\bar\psi(z;\tau)]\rho(z;\tau)}$ is the phase in the left
well while $\theta_R(\tau) =
\mathrm{arctan}\frac{\int_0^{\infty}\mathrm{d}z \mathrm{Im}
[\bar\psi(z;\tau)]\rho(z;\tau)} {\int_0^{\infty}\mathrm{d}z
\mathrm{Re}[\bar\psi(z;\tau)]\rho(z;\tau)}$ in the right well with
the density $\rho(z;\tau)= \bar\psi^*(z;\tau)\bar\psi(z;\tau)$.

$\phi_+(z)$ and $\phi_-(z)$ represent the initial ground state and
the first excited state wave functions for the condensates in the
double-well. Their linear combinations can be defined as the left
(right) mode: $\psi_{L,R}(z)=\frac{\phi_+(z)\pm\phi_-(z)}{2}$. They
satisfy the orthogonal condition
$\int\mathrm{d}z\psi_L(z)\psi_R(z)=0$. The trial wave function for
obtaining the initial state can be chosen as the superposition of
$\psi_L(z)$ and $\psi_R(z)$ as in Ref. \protect\cite{2M}
$\bar\psi(z;\tau)=\psi_L(\tau)\phi_L(z)+\psi_R(\tau)\phi_R(z)$,
where $\psi_{L(R)}(\tau) =
\sqrt{n_{L(R)}(\tau)}\mathrm{e}^{i\theta_{L(R)}(\tau)}$. The
population imbalance and relative phase at time $\tau$ are defined
as $\Delta n=n_L-n_R$ and $\Delta\theta=\theta_L-\theta_R$. A given
trial at $\tau=0$ is
$\bar\psi(z;0)=\mathrm{e}^{i\Delta\theta(0)}\sqrt{n_L(0)}\psi_L(z)+\sqrt{n_R(0)}\psi_R(z)$.
$\Delta n(0)=n_L(0)-n_R(0)$ represents the initial population
imbalance and $\Delta\theta (0)$ the phase difference of condensates
in double well. The relative phase can be measured by the interfere
patterns of releasing the Bose-Einstein condensates from the
double-well potential after different evolution times
\cite{Oberthaler2005}. Moreover, a technique based on stimulated
light scattering has been developed to detect $\Delta\theta$
nondestructively \cite{Saba2005}.

The reduced time-dependent GP equation can be solved using the
Split-Step Crank-Nicolson scheme, with both the space and time being
discretized \cite{Adhikari}. In our calculation, time step is
$\delta \tau=0.001$ and space step is $\delta z=0.01$. The initial
ground state and the first excited state wave functions, $\phi_+(z)$
and $\phi_-(z)$, can be obtained by the imaginary-time propagation.
The dynamical evolution are calculated by solving Eq. (\ref{GP1D})
using real-time propagation method.

\begin{figure}[tb]
\center\includegraphics[width=0.8\linewidth]{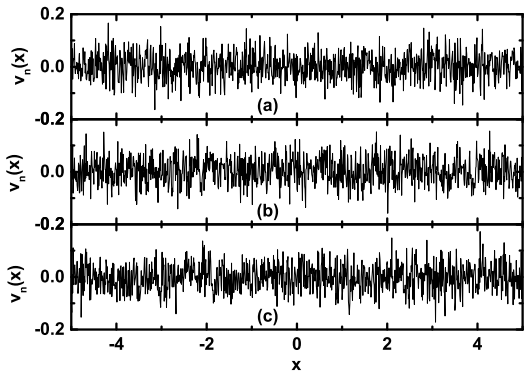}
\caption{Snapshots of the random Gaussian white noise potential with
the noise strength $D=0.005$ at the time $\tau=1$ (a), $\tau=2$ (b),
and $\tau=10$ (c), respectively. } \label{fig:N}
\end{figure}

The Gaussian white noise is produced numerically by the Box-Mueller
algorithm \cite{Ronald1988},
\begin{eqnarray}\label{vn}
v_n=\sqrt{-4D\ln (a)}\cos(2\pi b),
\end{eqnarray}
where $a$ and $b$ are two uniformly distributed random
numbers on an unit internal, and the dimensionless parameter
$D=D_0/({\hbar\omega_z})^2$. $v_n$ varies occasionally from time to
time. Figure~\ref{fig:N} displays snapshots of the spatial form of
the random potential at different given time. Suppose that it
changes $R$ times each unit time. Then $R$ defines a character
parameter with respect to the velocity of noise and it is called the
change rate of noise hereinafter. The fastest noise is the one whose
step interval $\Delta t$ just amounts to the calculation time step
$\delta t$. It means that the noise potential changes once per
calculation step. Since the time step $\delta \tau=0.001$, the
change rate of the fastest noise in our study is $R=1000$.

%The TDGP equation we used in the following only has one-space-variable now. It is a partial differential equation in space and time variables including first-order time and second-
%order space derivatives. A commonly accepted semi-implicit Crank-Nicolson discretization scheme is used here to find the solution of the equation.
%Space and time are discretized and subsequently the discretization equation is integrated and propagated. The numerical algorithm is executed in two parts.

%%%%%%%%%%%%%%%%%%%%%%%%%%%%%%%%%%%%%%%%%%%%%%%%%%%%%%%%%%%%%%%%%%
\section{Results and discussions}
\label{sect:diagram}

%\begin{figure}[htb]
%\includegraphics[width=0.95\linewidth]{Noise.eps}
%\caption{Illustration of the double-well trap with Gaussian white noise. The amplification of noise is $D=100$.
%} \label{fig:trap}
%\end{figure}

\begin{figure}[tb]
\center\includegraphics[width=0.8\linewidth]{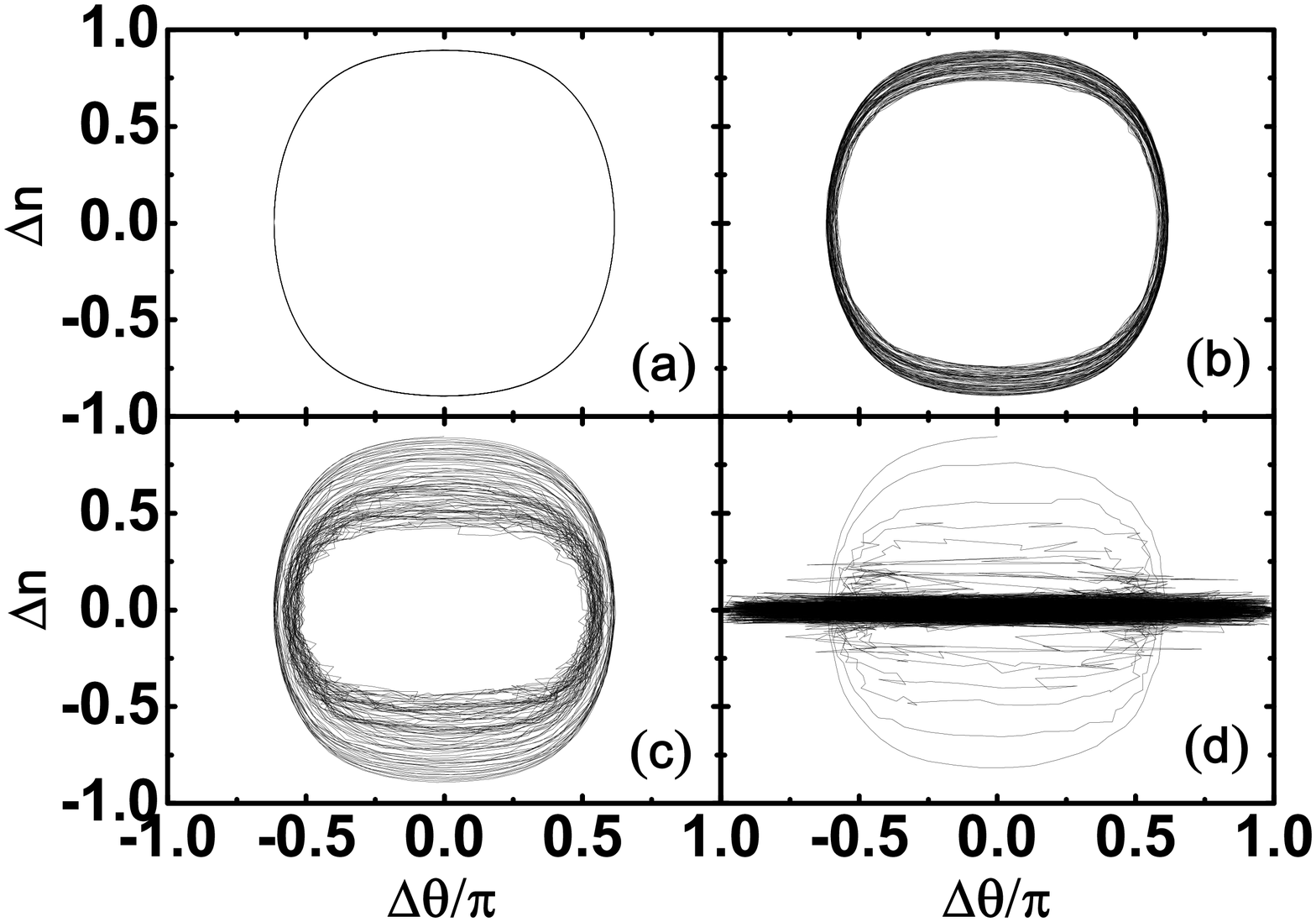}
\caption{Evolution trajectories in $\Delta\theta$-$\Delta n$ phase
space of a system initially in the Josephson oscillation state
($\Delta n=0.9$, $\Delta\theta=0$) with the fixed change rate of
noise, $R=1000$. The evolution time is from $\tau=0$ to $\tau=5000$
and the noise strength is $D=0$ (a), $D=0.005$ (b), $D=0.02$ (c),
and $D=0.5$ (d), respectively. The black horizontal band in panel d
consists of many almost vertical lines and it denotes the
irregularly energy-fluctuating states in the final evolution stage.
} \label{fig:JO}
\end{figure}

The phase-space diagram of double-well BECs without noise has been
studied based on the time-dependent GP equation \cite{gu12} and the
obtained results are consistent quantitatively with the two-mode
model results \cite{Smerzi1997}. Three typical dynamical regimes can
be present. (i) The Josephson oscillation regime consists of a
cluster of close orbits circling the lowest energy point $(\Delta
n=0,\Delta\theta=0)$ or $(\Delta n=0,\Delta\theta=2k\pi)$ where $k$
is an integer. (ii) The $\pi$-mode MQST regime consists of a cluster
of close orbits circling the highest energy point $(\Delta n\ne
0,\Delta\theta=(2k\pm 1)\pi)$. (iii) The running-phase MQST regime
consists of open orbits which lie between the above two regimes.

To obtain the numerical results of dynamical double-well condensate
subject to the Gaussian white noise, we set the double-well
potential in Eq. (\ref{GP1D}) to be
$v_{dw}(z)=z^2+10\mathrm{exp}(-z^2)$ for simplicity. The interatomic
interaction $g_{1D}$=$0.01$. Under above choice of relevant
parameters, all three dynamical regimes appear in the phase diagram.
First, we reproduce the phase diagram without noise, $D=0$, and
determine the location of each typical dynamical regime. Then we
choose one initial state from each regime by setting the initial
particle imbalance $\Delta n$ and relative phase $\Delta\theta$
appropriately.

%Translating the intensity of the noise $D$ to the potential language, we have the amplitude of the fluctuation of the potential randomly ranging from $-0.1\sqrt{D}$ to
%$0.1\sqrt{D}$. (???abolish???)

%What we should emphasis here is that we only consider the effects of the Gaussian white noise, the dissipation effects are not included.

\begin{figure}[tb]
\center\includegraphics[width=0.8\linewidth]{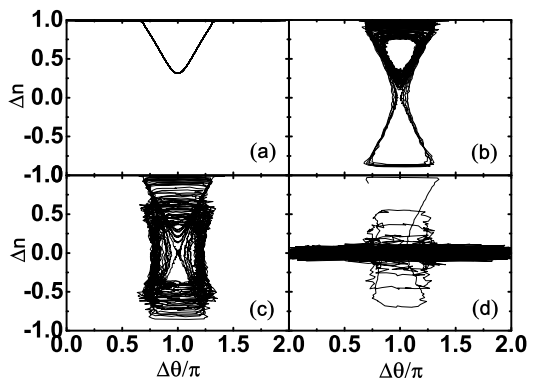}
\caption{Evolution trajectories in $\Delta\theta$-$\Delta n$ phase
space of a system initially in the running phase MQST state ($\Delta
n=0.9$, $\Delta\theta=0.7\pi$) with the fixed change rate of noise,
$R=1000$. The evolution time is from $\tau=0$ to $\tau=5000$ and the
noise strength is $D=0$ (a), $D=0.005$ (b), $D=0.02$ (c), and
$D=0.5$ (d), respectively. The black horizontal band in panel d
describes the same evolution stage as in Fig.~\ref{fig:JO}(d). }
\label{fig:Running-MQST}
\end{figure}

We first consider the dynamical evolution subject to Gaussian white
noise with high change rate of $R=1000$. Fig.~\ref{fig:JO} (a) shows
the orbits presented by constant energy lines for an initial state
of Josephson oscillation ($\Delta n$=$0.9$, $\Delta\theta$=$0$)
without noise ($D=0$). Both the population imbalance and the
relative phase oscillate around the zero point ($\Delta\theta=0$,
$\Delta n=0$ ) during the evolution. When Gaussian white noise with
the strength $D=0.005$ is imposed on, as illustrated in
Fig.~\ref{fig:JO}(b), the trajectory gets a little ``fat'', but
dynamical properties are not qualitatively different from the system
without noise. However, the noise effect shows up apparently when
$D=0.02$, as seen in Fig.~\ref{fig:JO}(c). The system evolves
gradually along an inward spiral path which is still smooth at
earlier stage but then becomes zigzag after a long time. During this
period, the system undergoes energy dissipation.
Fig.~\ref{fig:JO}(d) shows that the energy dissipates more quickly
as the noise becomes extremely stronger ($D=0.5$). At last, the
original Josephson oscillation has been completely destroyed and the
system enters into a state in the black horizontal band along the
$\Delta n=0$ line indicated in Fig.~\ref{fig:JO}(d). The particle
imbalance $\Delta n$ is almost zero which means that particles tend
to distribute almost equally in each well, while the relative phase
varies very fast, signaling that $\Delta\theta$ is no longer a
well-defined parameter. It is also worth noting that the energy does
not decrease monotonically any longer, but becomes irregularly
fluctuating. Therefore, such a state is called the irregularly
energy-fluctuating state in this paper. And it is also the final
state of the system after a long enough evolution, no matter how
weak the noise strength is.

\begin{figure}[tb]
\center\includegraphics[width=0.8\linewidth]{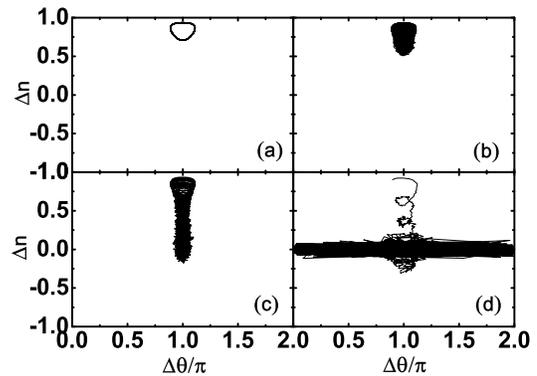}
\caption{Evolution trajectories in $\Delta\theta$-$\Delta n$ phase
space of a system initially in the $\pi$-mode MQST state ($\Delta
n=0.9$, $\Delta\theta=0.9\pi$) with the fixed change rate of noise,
$R=1000$. The evolution time is from $\tau=0$ to $\tau=5000$ and the
noise strength is $D=0$ (a), $D=0.005$ (b), $D=0.02$ (c), and
$D=0.5$ (d), respectively. The black horizontal band in panel d
describes the same evolution stage as in Fig.~\ref{fig:JO}(d).}
\label{fig:MQST}
\end{figure}

Fig.~\ref{fig:Running-MQST} and \ref{fig:MQST} demonstrate
situations of the initial state being in the running phase MQST
regime ($\Delta n$=$0.9$, $\Delta\theta$=$0.7\pi$) and $\pi$-mode
MQST ($\Delta n$=$0.9$, $\Delta\theta$=$0.9\pi$), respectively. In
this case, each MQST regime splits into two separate parts which are
symmetric with respect to the $\Delta n=0$ line in the phase space.

Look at the running phase MQST case. When the system is subject to a
weak noise, the phase may run for a few periods and then move around
a energy maximum point whose evolution trajectory looks like the
$\pi$-mode MQST. If the noise strength turns stronger, the system
starts $\pi$-mode-like oscillating from very beginning, as shown in
Fig.~\ref{fig:Running-MQST}(b). One interesting phenomenon is that
the trajectory may pass across the $\Delta n=0$ line and then
evolves around the down energy-maximum point $(\Delta n<0,
\Delta\theta=\pi)$. This process will be accelerated if $D$ is
increased and the system may be in a Josephson-like oscillation
around the point ($\Delta n$=$0$, $\Delta\theta$=$\pi$), as shown in
Fig.~\ref{fig:Running-MQST}(c). The energy decreases with time in an
oscillatory manner. In case of extremely strong noise, the system
falls down to the irregularly energy-fluctuating state after a short
period of damping.

Fig.~\ref{fig:MQST} displays the cases evolving initially from the
$\pi$-mode MQST state ($\Delta n$=$0.9$, $\Delta\theta$=$0.9\pi$).
It is obvious that the $\pi$-mode MQST state is more sensitive to
the noise. As shown in Fig.~\ref{fig:MQST}(b), with the
participation of noise as weak as $D=0.005$, the previous closed orbit
has already been becoming very fuzzy. When the noise strength is
increased to $D=0.02$, the system gets into running phase regime
quickly. The following evolution can be regarded as a process
starting from a running phase MQST state. The basic features should
be similar to Fig.~\ref{fig:Running-MQST}. The actual evolution
trajectory depends on the concrete initial state. This is the reason
why Fig.~\ref{fig:MQST}(c) and \ref{fig:MQST}(d) look quite
different from Fig.~\ref{fig:Running-MQST}(c) and
\ref{fig:Running-MQST}(d).

\begin{figure}[tb]
\center\includegraphics[width=0.8\linewidth]{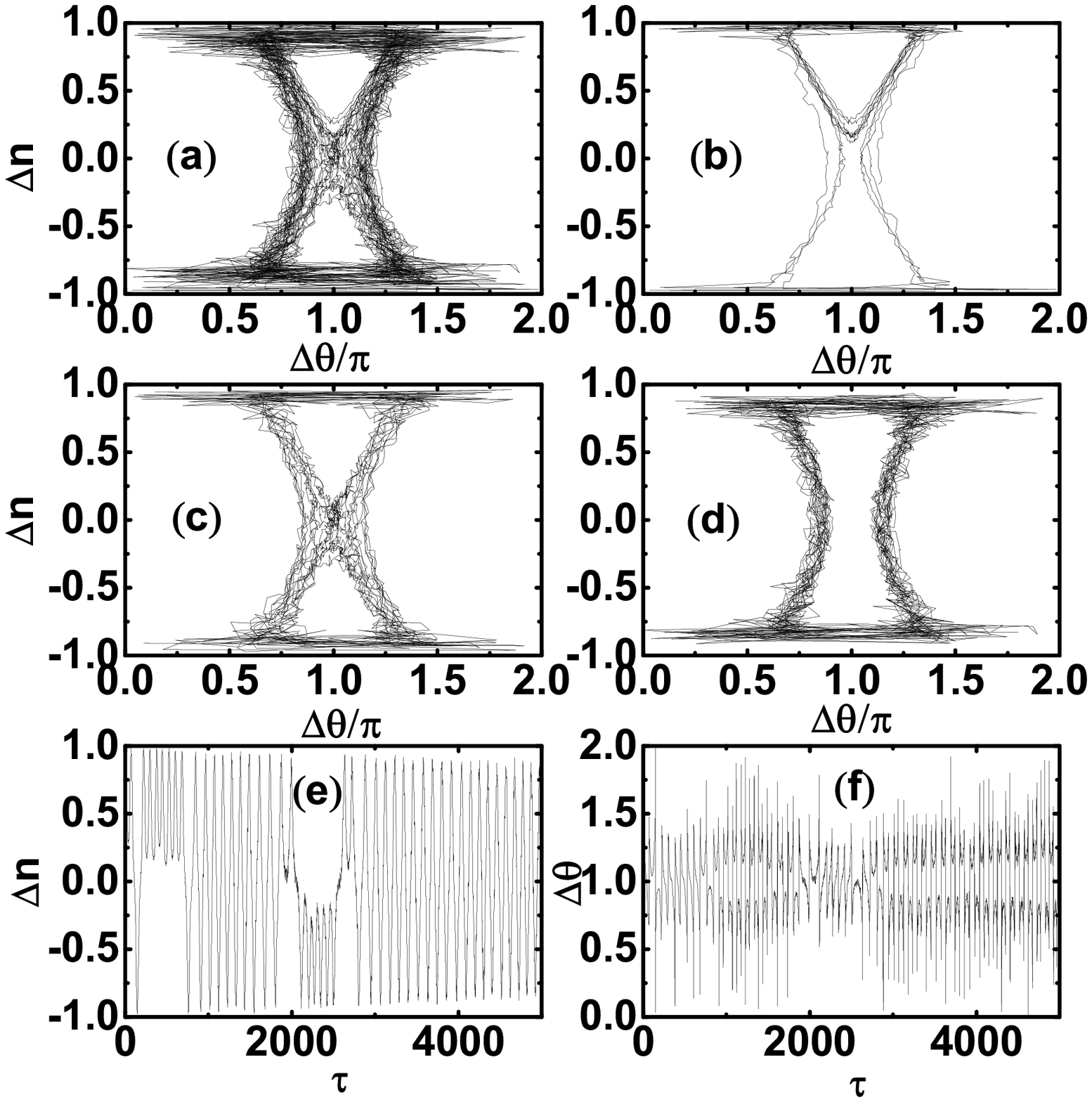}
\caption{Evolutions of an initially running phase MQST state
($\Delta n=0.9$, $\Delta\theta=0.7\pi$) in presence of noise with
the noise strength $D=0.005$ and the change rate $R=10$. The
trajectory is plotted for the evolution time from $\tau=0$ to
$\tau=5000$ (a), from $\tau=0$ to $\tau=1000$ (b), from $\tau=1000$
to $\tau=3000$ (c), and from $\tau=3000$ to $\tau=5000$ (d),
respectively. The $\Delta n$-$\tau$ plot (e) and
$\Delta\theta$-$\tau$ plot (f) are shown in the third row. }
\label{fig:R}
\end{figure}

According to discussions above, when the noise strength is small,
the Gaussian white noise may lead to an energy dissipation effect.
The noise with a larger $D$ makes the energy dissipating faster. But
there are some differences in the dissipative process between the
present study and the results obtained within phenomenological
theories \cite{Marino,gu12}. The main difference lies in the
evolutions from MQST states. The system damps directly towards the
saddle point ($\Delta n$=$0$, $\Delta\theta$=$\pi$), not the energy
minimum ($\Delta n$=$0$, $\Delta\theta$=$0$). Then it steps into the
irregularly energy-fluctuating state. The phenomenological theories
suggest that the energy decreases all the time until the system
approaches the energy minimum point. No matter how strong the
dissipation is, the system will finally damp into the energy minimum
state, then the evolution stops. The irregularly energy-fluctuating
state does not appear.

We have calculated the dynamical behaviors of the double-well
condensate subject to Gaussian white noise with different strength
$D$ and change rate $R$. The noise effect is mainly determined by
the strength $D$ when the change rate $R$ is large enough.
Nevertheless, the change rate $R$ also plays an important role in
understanding the dynamical evolution when $R$ is small.

Figure~\ref{fig:R} displays a typical case of the noise with small
strength ($D=0.005$) and slow change rate ($R=10$). The evolution is
from an initial state of running phase MQST ($\Delta n$=$0.9$,
$\Delta\theta$=$0.7\pi$). Figure~\ref{fig:R}(b) looks like
Fig.~\ref{fig:Running-MQST}(b), but the evolution time is only
within $\tau=1000$, far shorter that $\tau=5000$ as in
Fig.~\ref{fig:Running-MQST}(b). Figure~\ref{fig:R}(d) shows the
trajectory with the time from $\tau=3000$ to $\tau=5000$. During
this period, the system oscillates in a Josephson-like way around
the saddle point ($\Delta n$=$0$, $\Delta\theta$=$\pi$), similar to
the case shown in Fig.~\ref{fig:Running-MQST}(c). It seems that
decreasing the change rate gives rise to similar effect as
increasing the noise strength. The $\Delta\theta$-$\tau$ and $\Delta
n$-$\tau$ plots shown in Fig.~\ref{fig:R}(e) and (f) provide
detailed information of the evolution process. Atoms can be trapped
almost in one well for a while and then oscillate between the two
wells. The variation range of $\Delta\theta$ is significantly
enlarged and many extremely sharp peaks appear in the
$\Delta\theta$-$\tau$ line. These peaks indicate that the relative
phase can change extremely fast, which may signal that the relative
phase is no longer well-defined, as discussed above. Such situation
occurs preferently in the large $\Delta n$ region, as indicated by
the almost vertical lines in Fig.~\ref{fig:R}(a). This is different
from the large $D$ case, in which the almost vertical lines appear
in the $\Delta n=0$ region first. If the system evolves further,
more almost vertical lines appear and they tend to cover the whole
phase space in an irregular manner. As a result, it is almost
impossible to expect in which state the system is at a given time.
Meantime, the energy of the system does not simply tend to damping,
but may go up and down occasionally. So the energy dissipation
effect seems smeared out during this stage.

%%%%%%%%%%%%%%%%%%%%%%%%%%%%%%%%%%%%%%%%%%%%%%%%%%%%%%%%%%%%%%%%%%
\section{Conclusion}\label{sect:conclusion}

In conclusion, we have investigated how an external Gaussian white
noise affects the dynamics of condensates in double-well. Dynamical
evolutions from three typical dynamical regimes of Josephson
oscillation, $\pi$-mode MQST and running phase MQST are discussed. It
is shown that the system keeps its original dynamical feature for a
long time when it is subject to a noise with weak
strength and rapid change rate which we called the weak noise. In
this case, the noise induced energy dissipation effect is observed.
Energy of the system decreases with evolution monotonously in the
Josephson oscillation regimes, and oscillatorily in the MQST
regimes. The difference between results of the present study and the
phenomenological theory is discussed.

Either increasing the noise strength $D$, or decreasing the noise
change rate $R$ may give rise to significant influence on dynamics
of the system. If $R$ is large enough, the influence of $D$ is quite
clear. The larger $D$ is, the faster the state evolves and damps.
Moreover, we also figure out that the noise with large strength
drives the system finally into the irregularly energy-fluctuating
state, in which the population imbalance tends to being very small
and the relative phase is no longer well-defined. This phenomenon is
different from the dissipation effect which makes the condensates
evolve from the high energy state to the lowest energy state where
the particle imbalance is zero and the relative phase is $2k\pi$
($k$ is an integer). Decreasing $R$ brings about similar effect as
increasing $D$ in the early stage of evolution. After a period of
time, the noise with slow change rate compels system to change
irregularly, with the relative phase being destroyed. But this
phenomenon happens preferently in the large $\Delta n$ region in
presence of the noise with slow change rate, which is different from
the effect caused by the noise with large noise strength but fast
change rate.

%And the final state of system is different as to different initial state such as Josephson oscillation state and MQST state. The original Josephson oscillation state finally becomes to $\pi$-mode MQST state while the original MQST (no matter the MQST is $\pi$-mode MQST or running phase MQST) finally turns into Josephson oscillation state. The relative phase has very small oscillation periods and the atoms tunnel through the barrier very quickly. But an important point we must highlight is that we have not found out the law of how the change rate of Gaussian white noise affects the system.
%that strong noise will shorten the evolving procedure to the final stage.

\begin{acknowledgments}
This work is supported by the National Natural Science Foundation of
China (Grant No. 11074021 and No. 11004007) and the Fundamental
Research Funds for the Central Universities of China.
\end{acknowledgments}

%%%%%%%%%%%%%%%%%%%%%%%%%%%%%%%%%%%%%%%%%%%%%%%%%%%%%%%%%%%%%%%%
%\section*{References}

\end{document}